\documentclass{prop2015}
\usepackage[english]{babel}

\keywords{Open quantum system, thermodynamics, Lindblad operator,
  quantum dynamics, harmonic chain.}
\title{Thermodynamic deficiencies of some simple Lindblad operators}
\subtitle{A diagnosis and a suggestion for a cure}
\author[J.\,T. Stockburger]{J\"{u}rgen T. Stockburger\inst{1}%
\footnote{Corresponding author\quad%
E-mail:~\textsf{juergen.stockburger@uni-ulm.de}}}
\author[T. Motz]{Thomas Motz\inst{1}}
\address[1]{Ulm University, Institute for Complex Quantum Systems}
\begin{abstract}
Master equations of Lindblad type have attained prominent status in
the fields of quantum optics and quantum information since they are
guaranteed to satisfy fundamental notions of quantum dynamics such as
complete positivity. When Lindblad operators are used to describe
thermal reservoirs in contact with an open quantum system, the
fundamental laws of thermodynamics and the fluctuation-dissipation
theorem provide additional mandatory criteria. We show several
examples of innocent-looking Lindblad operators which have
questionable properties in this regard. Compatibility criteria between
Hamiltonian and Lindblad terms as well as consequences of their
violation are discussed. An alternative stochastic approach to
dissipative quantum dynamics is outlined and illustrated through a
harmonic-chain model for which the approach of local Lindblad
operators fails.
\end{abstract}
\shortabstract
\newcommand{\tr}{\mathop{\mbox{tr}}}
\newcommand{\var}{\mathop{\mbox{var}}}
\renewcommand{\Re}{\mathop{\mbox{Re}}}
\renewcommand{\Im}{\mathop{\mbox{Im}}}
\begin{document}
\let\otoday\today
\renewcommand{\today}{LaTeX formatted \otoday}
\maketitle

\section{Introduction}
There has been a recent growth of research activity in the study of
the thermodynamics of small (mesoscopic or microscopic) quantum
systems. A major part of these activities brings the fields of
thermodynamics and dynamics into contact, with interesting results
beyond the standard approaches of small fluctuations and linear
response.

The smaller a system is, the less isolated it tends to be, hence the
concept of an open quantum systems is frequently referred to in
current efforts, and established methodology related to this concept
is typically used. One of the premier mathematical tools in this
context is the Lindblad master equation. Its ubiquitous use in the
literature, however, should not be taken as evidence of universal
applicability. Levy and Kosloff \cite{levy14} have put forward
strong evidence to the contrary, showing that certain ways of
constructing local Lindblad operators violate thermodynamics.

A few words on their epistemological status and use in the context of
thermal physics seem therefore in order. There are two conceptually
different approaches to open quantum systems which overlap to some
degree, but not completely. On the one hand, the notion of an open
quantum system may be introduced in an abstract manner, starting from
the question how the Liouville-von Neumann equation can be generalized
to non-unitary, dissipative time evolution. When the requirements of
unitarity and energy conservation are dropped, the remaining
requirements are taken to be linearity, complete positivity, and the
existence of a superoperator which acts as a generator of the
dynamics. Starting from these requirements, one arrives at quantum
semigroups whose generator can be decomposed into Hamiltonian and
Lindblad terms \cite{lindb76,alick87}. The physical context in which
this methodology has found its most frequent application is quantum
optics.

A more physically motivated approach, which may also give rise to
Lindblad-type (and other) master equations, relies on the established
postulates of (conservative) quantum mechanics only. The physical
environment towards which the system of interest is ``open'' must then
be part, at least conceptually, in the starting point of such an
approach. Using a sequence of approximations \cite{breue02},
    Lindblad operators can be identified as modifications of the
    system dynamics by the reservoir, after a partial trace is taken
    over all environmental degrees of freedom.

In this procedure, the Born approximation for system-environment
interaction is assumed for timescales of the order of decoherence and
relaxation timescales, and a further weak-coupling assumption is
implied in the secular approximation which is needed to bring the
resulting master equation into Lindblad form.

A few comments seem in order here: Strict application of the
methodology just outlined requires some knowledge of the spectral
power density of reservoir fluctuations, and the diagonalization of
the system Hamiltonian (at least within an uncertainty given by
Planck's constant divided by the correlation time of reservoir
fluctuations). For some applications, techniques reaching beyond the
Born approximation are needed, either using higher-order corrections
\cite{shiba77} or entirely different approaches for reducing the
dynamics of system-plus-reservoir models, e.\,g., path integrals in
the case of quantum Brownian motion \cite{grabe88} or for the
instanton theory of dissipative tunneling \cite{weiss87a}.  Stochastic
modeling techniques \cite{stock99,strun99,stock02}, equivalent to
non-perturbative path integral approaches, are sometimes attractive as
computational methods.

\section{Inherent limits of the standard Lindblad approach in
  thermodynamic contexts}\label{sec:inhlim}

In a thermodynamic context, the system-plus-reservoir approach is
the appropriate reference point; heat baths are then identified as
``environment''. Under certain conditions, the standard Lindblad master
    equation
\begin{equation}
\label{eq:Lindblad}
{d \rho\over dt} = -i [H,\rho]
+  \sum_j L_j \rho L_j^\dagger - {\textstyle{1\over 2}}
\{L_j^\dagger L_j, \rho\}
\end{equation}
can be derived for this case \cite{breue02,hbar}. In this
thermodynamic setting, it acquires the following additional
properties:
\begin{enumerate}
\item The dynamics of diagonal elements of the density matrix
  (relaxation) decouples from the dynamics of the off-diagonal
  elements (coherent evolution and dephasing).
\item Transition rates between the diagonal elements obey a detailed
  balance condition, determined by the level structure and the
  reservoir temperature.
\item The Gibbs ensemble related to the system Hamiltonian is a
  stationary state of the master equation. In the absence of special
  symmetries the stationary state is unique; the master equation
  describes thermalization.
\item When the Hamiltonian and dissipative parts of the Liouvillian
  are considered separately, the Gibbs state is a zero-eigenvalue
  eigenstate of either part.
\end{enumerate}
From these points, it is evident, that the \emph{unitary dynamics} of
the system-plus-reservoir model can be used to describe
thermalization of the system, at least under the weak-coupling
provisos on which the derivation of the Lindblad master equation
relies on in this context.

The stationary state of a Lindblad equation properly derived from a
system-plus-reservoir model is virtually always a valid thermodynamic
state, however, this does not allow the conclusion that it faithfully
reflects the true state of a microscopic system in finite-strength
contact with a heat bath. The stationary state basically results from
solving the detailed-balance conditions for the stationary
probabilities (and normalization). It lies in the nature of this
procedure that the result is of lower-order accuracy when
perturbatively determined rates are used. Rates are correct to second
order in the system-reservoir coupling here, resulting in the
zero-order Gibbs result.

It is to be noted that the zero-order result may be qualitatively
wrong in the case of low temperatures. The zero-order result depends
only on the system Hamiltonian. In the case of a gapped Hamiltonian,
and for thermal energies sufficiently below the gap, the zero-order
result show an essentially frozen system with exponentially small
thermodynamic quantities, e.g., a heat capacity $\propto
\exp(-E_1/k_{\rm B} T)$ (Schottky anomaly). For simple systems such
as the spin-boson model or the damped harmonic oscillator, the reduced
density matrix may be computed with better accuracy in this regime,
e.g, by path integral methods. The essential thermodynamic properties
at low temperature are actually related to the structure of a
hybridized, broadened system ground state, with thermodynamic
quantities varying algebraically, sometimes linearly with temperature
\cite{extensive}. This has been demonstrated for a two-state system
\cite{gorli88} and for the damped harmonic oscillator \cite{grabe88}.
For the generic model of Ohmic dissipation, both works find a heat
capacity $\propto T$. Ankerhold and Pekola have recently discussed
the experimental implications of such hybridization effects
\cite{anker14}.

Closely related to the path integral approach is the dynamical
ansatz of a hierarchy of equations of motion \cite{tanim90}
(augmenting the dynamical state by auxiliary density matrices). This
approach is useful when studying systems or physical effects where
system-reservoir correlations play a significant role, e.g., in
multi-dimensional spectroscopy.

\section{Heuristic Lindblad operators: When are they consistent with
  thermodynamics?}\label{sec:heur}

The derivation of the Lindblad master equation as a reduced dynamics
of a system-plus-reservoir model as referred to above can be a
formidable task. Not only must the system Hamiltonian be diagonalized,
the system-reservoir coupling must be decomposed into a linear
combination of raising and lowering operators between the energy
eigenstates of the system. In simple cases such as the harmonic
oscillator or noninteracting spin systems, the raising and lowering
operators are few and easy to determine. Complex systems with $n$
non-equidistant energy levels require a total of up to $n(n-1)$ such
eigenoperators of the Hamiltonian \cite{breue02}.

Because of this complexity, heuristic Lindblad master equations are
frequently postulated by ``defining'' an open system either by
modifying the Hamiltonian \emph{after} an ``orthodox'' Lindblad master
equation had been derived for a simpler, unperturbed Hamiltonian, or
by selecting ad-hoc combinations of Hamiltonian and Lindblad terms
by hand.

A most simple test case demonstrates problems with the former
approach (i.e., modifying the Hamiltonian after introducing a Lindblad
dissipator): Let us consider the standard Lindblad master equation for
the damped harmonic oscillator in a zero-temperature environment,
\begin{equation}
\label{eq:osclindstd}
\dot{\rho} = -i\omega [a^\dagger a,\rho] + \gamma a\rho a^\dagger
-\textstyle{\frac{\gamma}{2}} \{a^\dagger a, \rho\}\;,
\end{equation}
which describes exponential decay to the ground state, but
modify the Hamiltonian by adding a linear force term,
\begin{equation}
H = H_0 + H_1 = \omega (a^\dagger a
 + \textstyle{\frac{1}{2}}) - f (a^\dagger + a)/\sqrt{2}\;.
\end{equation}
Properties of the stationary state of this dynamics can be
conveniently determined by considering the adjoint (Heisenberg) master
equation for an observable $A(t)$, \cite{breue02,adjoint}
\begin{equation}
\label{eq:adjlindblad}
\dot{A} = - i\omega [A,H] + \gamma a^\dagger A a
-\textstyle{\frac{\gamma}{2}} \{a^\dagger a, A\}\;.
\end{equation}
The \emph{stationary} solutions for position $q =
(a+a^\dagger)/\sqrt{2}$ and momentum $p = i (a^\dagger-a)/\sqrt{2}$
are (to leading order in $\gamma$)
\begin{equation}
q_{\rm stat} = \frac{f}{\omega},\quad p_{\rm stat} = - \frac{\gamma f}{2
  \omega^2}\neq 0\;.
\end{equation}
The former equation is Hooke's law in the mass-reduced units used
here; the latter equation is an unphysical result (see also
\cite{grabe06}).

Qualitative changes to the Hamiltonian can induce larger discrepancies
between thermodynamic states and the stationary state of the resulting
Lindblad master equation. Keeping the Lindblad operators of
eq. (\ref{eq:osclindstd}), but changing the Hamiltonian to that of a
free particle, $H = \frac{\omega}{4}p^2$ in the units used here, one
finds a stationary state which is a highly squeezed Gaussian with
\begin{eqnarray}
\langle p^2\rangle &=& \frac{1}{2}\\
\textstyle{\frac{1}{2}} \langle pq+qp\rangle &=& \frac{\omega}{2\gamma}\\
\langle q^2\rangle &=& \frac{\omega^2}{\gamma^2} + \frac{1}{2}\;.
\end{eqnarray}
This cannot be identified as a thermal state for several reasons. In a
non-magnetic model, there can be no correlations between $q$ and $p$.
Moreover, in a proper treatment (see, e.g., \cite{weiss12}), one finds
that there is no stationary state: By changing the Hamiltonian to that
of a free particle, we have effectively changed the problem to that of
quantum Brownian motion, which has no stationary state, $\langle
q^2\rangle$ diverges linearly with time.

Of course, not all changes to the Hamiltonian are this harmful. When a
Lindblad master equation is derived from a system-reservoir model, one
critical step is a decomposition of the system-reservoir coupling into
eigenoperators of the Hamiltonian \cite{breue02}. Starting from an
interaction term $H_{\rm I} = - A\cdot B$ which is separable between a
system part $A$ and a reservoir part $B$, Lindblad operators $L_j$ are
chosen such that $A$ is a linear combination of them, and with the
special property that the time dependence of $L_j(t)$ in the
interaction picture is
\begin{equation}
\label{eq:intdyneigenop}
L_j(t) = e^{i\omega_j t} L_j(0)\;.
\end{equation}
Here $\omega_j$ is one of the transition frequencies of the
system. Keeping the $L_j$ while changing the Hamiltonian will, of
course, lead to a different, non-trivial equation of motion of $L_j(t)$
in the interaction picture. However, as long as
eq. (\ref{eq:intdyneigenop}) remains a good approximation for the
duration of the \emph{reservoir correlation time}, only a small error
is incurred. In this context, it should not be overlooked that even
broad-band, unstructured reservoirs have a timescale which is not
always short in a quantum context: The thermal timescale $1/(k_{\rm
  B}T)$ is relatively long if $k_{\rm B}T$ is small compared to a
typical level spacing of the system. In essence, this means that the
modification of a Hamiltonian in a Lindblad master equation is safe if
it does not change the eigenstates appreciably, and if the level
shifts and splittings it induces are smaller than $k_{\rm B}T$.

In the light of these considerations, models defining Lindblad
operators defined entirely \emph{ad hoc} should be considered as more
or less disconnected from thermodynamics and statistical physics
unless cogent arguments are found for their validity.

Levy and Kosloff analyze such a case in a recent paper \cite{levy14},
where they study a ``dimer'' of harmonic oscillators (or,
alternatively, two-level systems) in order to test the compatibility
of \emph{site-local} Lindblad operators with thermodynamics. The local
dissipation operators considered there are a correct description for a
\emph{monomer} (in the sense of properties 1--4 enumerated in section
\ref{sec:inhlim}). Levy and Kosloff choose different temperatures for
the reservoirs of two monomers, then study the dynamics which arises
when the two monomers are coupled while leaving the site-local
Lindblad operators unchanged. For this case, they find significant
parameter regimes where the resulting dynamics has a nonequilibrium
steady state in fundamental disagreement with thermodynamics: Energy
is transferred from the cold to the hot side of the dimer without any
work being performed on the system.

Here we show that even the equilibrium state of such a model is
sometimes unphysical, even in highly symmetric cases (in particular,
reservoirs with identical temperatures). We consider a short chain of
three harmonic oscillators with Hamiltonian
\begin{equation}
H = H_0 + V_{\rm int} = \sum_{i=1}^3 \textstyle{\frac{1}{2}}
m_i \omega_i^2 q_i^2 + V_{\rm int}
\end{equation}
with interaction potential
\begin{equation}
V_{\rm int} = - \sum_{i=1}^2 \mu_i q_i q_{i+1}
\end{equation}
\begin{figure}
\includegraphics*[width=\columnwidth]{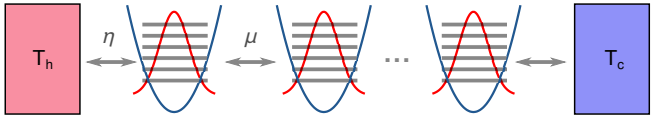}
\caption{Chain of harmonic oscillators coupled to Ohmic heat baths.}
\label{fig:oscchain}
\end{figure}
in the symmetric configuration $m_i = \omega_i = 1$, $\mu_i =
0.7$. The first and last oscillator is subject to the zero-temparature
Lindblad terms of eq. (\ref{eq:osclindstd}).

The stationary state of the resulting master equation, obtained using
an analogue of eq. (\ref{eq:adjlindblad}), differs significantly from
the ground state of the chain. These artifacts persist down to the
regime of extremely weak damping. For values of $\gamma$ between
$10^{-3}$ and $10^{-12}$, we find essentially the same Gaussian state
with a covariance matrix incompatible with ground state properties. It
appears that however small $\gamma$ is chosen, the same artifact
arises. Most likely this can be attributed to the fact that a
transformation of the annihilation operators $a_1$ and $a_3$ to a
normal-mode picture will also contain creation operators of the normal
modes. In a sense, the local Lindblad dissipator, however weak,
``pulls'' the system in the wrong direction.

The symplectic eigenvalues of the covariance matrix provide a
convenient route for the determination of the entropy of the resulting
state \cite{weedb12}. We find it is a mixed state with entropy
$S \approx 4.10 k_{\rm B}$. With $S \approx k_{\rm B} \ln n$, this
corresponds to roughly 60 occupied states.

Larger chains, such as those currently studied experimentally as
cold-atom replicas of solid-state systems, present problems even
outside the specific context of site-local damping operators: In the
limit of long chain length, energy levels come arbitrarily close to
each other. For large enough systems, the combination of the Born and
rotating-wave approximations inherent in the standard derivation of
Lindblad master equation \cite{breue02} leads to difficulties: The
assumption of a small level broadening compared to the level spacing
breaks down.  In any gapless model, low-lying single-particle
excitations will exhibit properties very similar to quantum Brownian
motion, leading to similar problems as those discussed earlier in this
context.

\section{Alternative formalism for open-system dynamics}

Some alternative formal approaches to system-plus-reservoir models,
such as path integrals, were already mentioned in the
introduction. Here we present recent work based on an exact stochastic
model of a harmonic or Gaussian quantum reservoir \cite{stock02}. The
key obstacle to the construction of such a model lies in the fact that
Heisenberg operators at different times generally do not commute; a
property which ordinary stochastic processes cannot reproduce.

Here is how to work around this problem:  Again starting from a
separable Hamiltonian $H_I = - A\cdot B$, it is to be noted that the
Liouvillian will be the sum of \emph{two} separable terms (interaction
picture):
\begin{equation}
\label{eq:dotrhoint}
\dot{\rho}(t) = i {A_+(t) B_-(t)} \rho(t) +i {A_-(t) B_+(t)} \rho(t)
\end{equation}
where superoperators for reservoir $B_{\pm}$, and system $A_{\pm}$
defined through (anti-)commutators, $B_- = [B,\cdot]$ and $B_+ =
\frac{1}{2} \{B,\cdot\}$ \cite{leftright}. It is now possible to
replace the \emph{superoperators} $B_+$ and $B_-$ in
eq. (\ref{eq:dotrhoint}) by c-number stochastic processes $\xi(t)$ and
$\nu(t)$ such that the stochastic mean value of solutions $\rho(t)$
will be exactly identical to the mean value obtained by tracing out
the quantum reservoir. Necessary and sufficient conditions for this
identification are
\begin{eqnarray}
\langle\xi(t)\xi(t')\rangle &=& \Re \langle B(t)B(t')\rangle\\
\langle\xi(t)\nu(t')\rangle &=& {2i}\Theta(t-t') \Im\langle
B(t)B(t')\rangle\\
\langle\nu(t)\nu(t')\rangle &=& 0\;.
\end{eqnarray}
These conditions may seem somewhat exotic for a ``classical''
stochastic process but can be fulfilled for complex-valued $\xi(t)$
and $\nu(t)$. For any correlation function $\langle B(t)B(t')\rangle$
whose real and imaginary parts are linked by the
fluctuation-dissipation theorem, this is a valid, exact model of a
quantum heat bath.

It is to be noted that the identification of stochastic average and
trace over reservoir is completely ``agnostic'' of the intrinsic
properties of the systems: The separable structure of the coupling is
preserved, and no decomposition of $A(t)$ into eigenoperators is
necessary. A set of processes $\xi(t)$ and $\nu(t)$ adapted to a
coupling Hamiltonian $H_{\rm I}$ and a reservoir correlation function
$\langle B(t)B(t')\rangle$ is valid for \emph{any} system Hamiltonian
$H_{\rm S}$. Virtually all problems outlined in the previous section
are thus avoided in this approach.

In the case of an Ohmic heat bath, the dissipative part $\Im\langle
B(t)B(t')\rangle$ is of the form $C \delta'(t-t')$ (derivative of
Dirac's delta function). This allows a time-local representation of
the dissipative part, while the fluctuation part $\Re \langle
B(t)B(t')\rangle$ becomes time-local only in the case of extremely
high temperatures \cite{calde83b}. The resulting equation of motion
\cite{stock99} then is, substituting the position operator $q$ for
$A$, and reverting to the Schr\"{o}dinger picture:
\begin{equation}
\label{eq:sled}
{d\over dt} \rho = -i
 \left( [H_{\mathrm{S}}, \rho] -
{\xi}(t) [q, \rho] \right)
 -i {\eta\over 2 m} [q,\{p, \rho\}]\;.
\end{equation}
The last term is identical to the friction term in the master equation
of Caldeira and Leggett \cite{calde83b}. However, explicit
fluctuations $\xi(t)$ are kept in eq. (\ref{eq:sled}), which is valid for
arbitrarily long thermal times $\beta$, i.e., arbitrarily low
temperature. Eq. (\ref{eq:sled}) is numerically stable and can used
for computation by direct sampling \cite{schmi11,schmi12,schmi13}.

We study harmonic chains of moderate size using deterministic
equations of motion based on eq. (\ref{eq:sled}). Also here, the
dynamics of the adjoint counterpart to the equation of motion
(\ref{eq:sled}) is useful, however, it is also a stochastic equation
of motion. All observables must be obtained through double (trace and
probabilistic) averages of the type
\begin{equation}
\langle A \rangle = \langle \tr(A\rho) \rangle_{\rm {prob}}\;.
\end{equation}
Variances related to this double average can be written as
\begin{equation}
\var (A,B) = \langle\, \rm {var}_{\rm {tr}} (A,B) \,\rangle_{\rm {prob}}
+ \rm {var}_{\rm {prob}} (\, \langle A \rangle_{\rm {tr}}  ,
\langle B \rangle_{\rm {tr}} \,)\;.
\end{equation}
The time evolution of $\rm {var}_{\rm {tr}} (A,B)$ can be obtained by
suitably combining the adjoint equations for $A$, $B$, and
$AB+BA$, which leads to the observation that  $\rm {var}_{\rm {tr}}
(A,B)$ evolves deterministically, independent of $\xi(t)$. This allows
the simplification
\begin{equation}
\var (A,B) = \rm {var}_{\rm {tr}} (A,B)
+ \rm {var}_{\rm {prob}} (\, \langle A \rangle_{\rm {tr}}  ,
\langle B \rangle_{\rm {tr}} \,)\;.
\end{equation}
The second term in the sum can be transformed using the observation
that $\langle A \rangle_{\rm {tr}}$ and $\langle B
\rangle_{\rm {tr}}$ are linked to $\xi(t)$ by linear response. The
phase-space vector $\vec{X}$ being of the form
\begin{equation}
\vec{X}(t) =  \vec{X}_0 + \int_0^t dt' \vec{G}(t-t') \xi(t')\;,
\end{equation}
the covariance matrix of position and momenta, or, more precisely, its
part contributed by terms of type $\rm {var}_{\rm {prob}} (\, \langle
A \rangle_{\rm {tr}}, \langle B \rangle_{\rm {tr}} \,)$ can therefore
be written as a matrix $\Sigma$ with
\begin{equation}
\Sigma(t) = \int_0^t dt' \int_0^t dt'' \vec{G}(t-t')
\vec{G}^\dagger(t-t'') \langle\xi(t')\xi(t'')\rangle\;.
\end{equation}
Using an auxiliary quantity $\vec{y}(t)$, a closed system of ordinary
differential equations can be solved to obtain $\Sigma(t)$:
\begin{eqnarray}
\dot\Sigma(t) &=& M \Sigma(t) + \Sigma(t) M^\dagger + \vec G(0) (\vec
y(t))^\dagger + \vec y(t) (\vec G(0))^\dagger \\
\dot{\vec{y}}(t) &=& \langle \xi(t)\xi(0)\rangle \vec G(t) \\
\dot{\vec{G}}(t) &=& M \vec G(t)\;.
\end{eqnarray}
The non-zero elements of the matrix $M$ consist of constants such as
masses, frequencies, couplings and friction constants. For a chain
coupled to reservoirs at either end, a slightly extended version of
these equations involving two processes $\xi_{\rm L}(t)$ and $\xi_{\rm
  R}(t)$ is needed.
\begin{figure}
\centering
\includegraphics*[width=0.85\columnwidth]{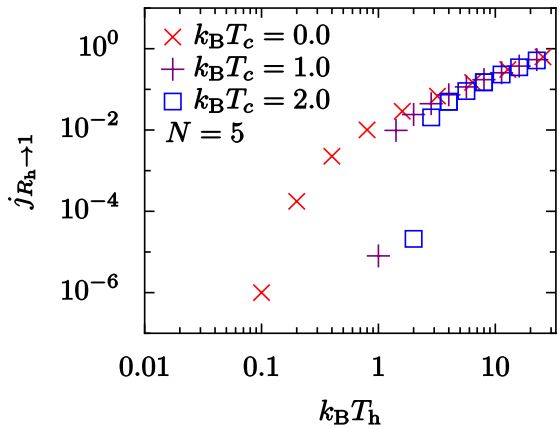}
\caption{Threshold behavior in the heat current through a harmonic
  chain of five oscillators. Parameters are $\mu=0.1$, $\eta=0.1$,
  reservoir high-frequency cut-off $\omega_c=30$.}
\label{fig:threshold}
\end{figure}
\begin{figure}
\centering
\includegraphics*[width=0.85\columnwidth]{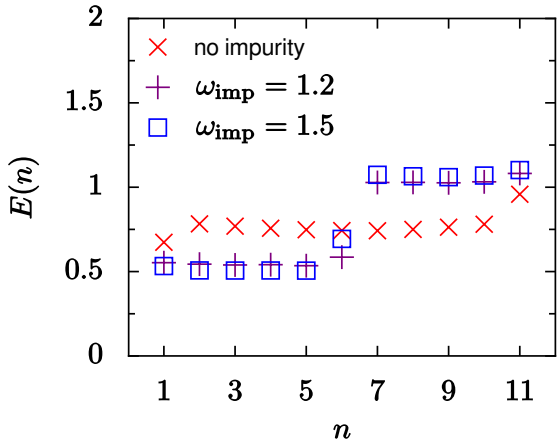}
\caption{Local oscillator energies for varying impurity strength.
Parameters are $\mu=0.1$, $\eta=0.1$,
$k_{\rm B}T_c = 0$, $k_{\rm B}T_h = 1$, 
  reservoir high-frequency cut-off $\omega_c=30$.}
\label{fig:impurity}
\end{figure}
Since these equations are now deterministic, both small and large
fluctuations can be determined with very moderate numerical
effort. The artifacts demonstrated for the local Lindblad approach --
unphysical heat currents and an unphysical equilibrium state -- are
completely absent. The direction of the heat current is determined
solely by the direction of the temperature gradient, and in the case
of equal reservoir temperatures we find the Gibbs state in the limit
of weak coupling.

Figure \ref{fig:threshold} shows heat currents through a chain of five
oscillators as a function of the temperature of the hotter of two
reservoirs, with different symbols designating different temperatures
$T_c$ of the colder reservoir. All three datasets show (in the log-log
representation chosen here) a threshold at which a positive current
sets in. If the temperature of the cold reservoir is finite, it sets
the scale for the threshold: The sign of the current correctly changes
with the sign of the temperature difference. A different (and maybe
more interesting) reason exists for the threshold in the third dataset
with $T_c=0$: In a chain of moderate size, the first excited state is
separated from the ground state by a small but finite gap. Transport
can occur only if one of the reservoir temperatures is sufficient for
thermal excitation of this state. Avoiding this finite-size effect
through larger chain lengths can be done with moderate effort: Our
dynamical variable $\Sigma(t)$ grows only quadratically with chain
length. Since our approach is non-perturbative, the breakdown of
perturbation theory for extended systems (outlined near the end of
section \ref{sec:heur}) does not occur here.

As another exercise, we make use of our liberty to change the
Hamiltonian without complications by introducing an impurity through
variation of the intrinsic frequency of one oscillator situated at the
center of an 11-oscillator chain (figure \ref{fig:impurity}). The
energy of each local oscillator is taken as an indicator for
differences in heat transport with and without impurity. Without
impurity, there is a fairly flat profile, indicative of ballistic
transport \cite{gaul07}. With an impurity, a step in this plateau is
introduced at the impurity, but otherwise, the profile remains
homogeneous.

\section{Conclusions and outlook}

There are several potential pitfalls when using Lindblad dissipators
to model thermal reservoirs. Some of them are intrinsic to the
approach, leading to subtle errors due to the perturbative nature of
their connection to system-plus-reservoir models. Further errors are
incurred when Hamiltonian terms and Lindblad operators do not match
in the sense outlined in sections \ref{sec:inhlim} and
\ref{sec:heur}. Qualitative errors may arise then, leading to
equlibrium or nonequlibrium states with properties which contradict
thermodynamics in a fundamental way.

In such cases, alternative descriptions of open-system quantum physics
are needed. These can be established methods like path integrals or
newer methods like hierarchic equation of motion or stochastic
modeling, both of which stay within the paradigm of an equation of
motion. The corresponding equations of motion define an essentially
non-perturbative approach. The computational cost of the stochastic
method in terms of explicit statistical sampling is not always
affordable, but whenever enough samples can be generated to achieve
numerical convergence, the result is guaranteed to be free of
systematic error.

We have analyzed certain aspects of heat transport through a harmonic
chain based on this methodology. For harmonic systems with ohmic
damping of arbitrary strength, deterministic equations of motion are
derived which are purely local (no diagonalization into eigenstates
or normal modes is required), yet we obtain the reduced system
dynamics of the underlying system-plus-reservoir model without the
errors described earlier. We are currently extending this approach to
larger chains, planning to investigate spatiotemporal patterns in
transport as well as the effects of disorder and/or weak
anharmonicity.

\subsection*{Acknowledgements}
We thank Joachim Ankerhold and Udo Seifert for stimulating
discussions. This work was supported by Deutsche Forschungsgemeinschaft
through grant AN336/6-1.




\providecommand{\noopsort}[1]{}
\providecommand{\WileyBibTextsc}{}
\let\textsc\WileyBibTextsc
\providecommand{\othercit}{}
\providecommand{\jr}[1]{#1}
\providecommand{\etal}{~et~al.}


\begin{thebibliography}{[10]}
\bibitem{hbar}%
We use the convention $\hbar=1$ throughout the paper.


\othercit
\bibitem{extensive}%
Thermodynamic quantities which are extensive in a macroscopic context are
  defined in this work for the system part by forming differences between the
  system-plus-reservoir complex and the unperturbed reservoir.
\bibitem{adjoint}%
Note that generally $(AB)(t) \neq A(t)B(t)$ for the non-unitary propagation of
  the adjoint master equation.
\bibitem{leftright}%
Alternatively, the two separable terms could be constructed from ``left'' and
  ``right'' superoperators.
\end{thebibliography}
\end{document}